\documentclass[%
 preprint,
 amsmath,amssymb,
 aps,
 longbibliography,
]{revtex4-1}

\usepackage{graphicx}
\usepackage{natbib}
\usepackage{amsmath}
\usepackage{bm}
\usepackage{subfiles}
\usepackage{color}
\usepackage{url}
\usepackage{cleveref}
\usepackage{epstopdf} 

\begin{document}
\title{On the Dynamics of a Quasi-Two-Dimensional Pulsed-Fludized Bed} 

\author{J.E. Higham}
 \email{j.e.higham@liverpool.ac.uk}
  \altaffiliation[Also at ]{Oakridge Institute for Science and Education, Oak Ridge, TN, USA}
\author{M. Shahnam}
\author{A. Vaidheeswaran}
  \altaffiliation[Also at ]{West Virginia University Research Corporation, Morgantown, WV, USA}

\date{\today}
\begin{abstract}
\noindent Periodic fluidization of solids in a gas medium can generate structured bubbling patterns. This phenomenon has been successfully reproduced in fluidized bed systems called pulsed-fluidized beds, known for their efficient mixing and tunability. The resulting pattern depends on particle properties, dimensions of the bed, mean inlet velocity, bed height, and amplitude and frequency of pulsing. It is however important to understand the changes in granular rheology associated with bubbling patterns. In the experimental work presented, we report on the results from a small-scale quasi-two-dimensional pulsed-fluidized bed. Proper Orthogonal Decomposition analysis applied to Particle Tracking Velocimetry data reveals a redistribution of energy between the harmonic and sub-harmonic modes as the inlet gas frequency is decreased. The change in bubbling pattern is accompanied by a change in granular dynamics due to the dominant modes of particle-phase motion.
\keywords{Proper Orthogonal Decomposition, pulsed-fluidized bed, bubbling bed, granular}
\end{abstract}

\maketitle

\section{Introduction}
\noindent Fluidization of solid particles by a gas medium has been studied in detail over the past few decades. The nature of fluidization is mainly dependent on the particle properties and operating conditions of the system \citep{Kunii1991}. Bubbles or regions of void are formed once the minimum fluidization velocity is reached. Dense bubbling fluidized beds are commonly used in several industrial applications including combustion, gasification, catalytic cracking and food processing \citep{Kunii1991}. The exchange of heat and mass between the phases is significant in these systems due to vigorous mixing of solids. This is a result of turbulence in the emulsion as bubbles rise. The particles are entrained in the wake of these bubbles  and are transported over significant distances in the stream-wise direction. The motion of particles is also impeded by inelastic interactions with the neighboring particles resulting in energy dissipation. The non-linear particle-particle and inter-phase interactions results in a chaotic system \citet{Fullmer2017}. 

Recent studies \citep{coppens2003structuring,massimilla1966study,bokun1968experimental,van2007four,Pence98,koksal1998bubble} have shown that the bubbling pattern in dense fluidized beds could be controlled using a periodic pulsation of gas flow at the inlet. \citet{massimilla1966study} observed that by increasing the frequency of the pulsing pattern, while maintaining a constant amplitude and baseline flow, the fluidized beds could be structured. Such pulsed-fluidized beds have been shown to be efficient in terms of mixing and heat transfer characteristics \citep{akhavan2008improved,li2004investigation,bokun1968experimental,ali2012fluidization,liu2003modeling}. The oscillatory flow rate at the inlet is proven to suppress chaos through phase-locking \citep{Pence98}. Hence, it is possible to control the mixing process which is challenging to achieve in conventional fluidized beds. However, research pertaining to these systems is primarily restricted to analyzing the effect of particle characteristics \citep{hernandez2011comparison,laverman2008investigation} and inlet gas properties \citep{coppens2003structuring,bizhaem2013experimental,beetstra2009influence,massimilla1966study} on the resulting flow structures. There exists a wide range of spatial and temporal scales making it difficult to formulate a framework to describe hydrodynamics \citep{kadanoff1999built} or energy cascade \citep{bocquet2001granular} across the different scales. Such efforts are limited and includes the work of \citet{laverman2008investigation} analyzing the vortical structures in these systems.

In the work presented, we report on the results from bench-scale pulsed-fluidized bed experiments performed at the U.S. Department of Energy's, National Energy Technology Laboratory (NETL). Proper Orthogonal Decomposition (POD) of flow field is used to elucidate complex granular dynamics at different operating frequencies while maintaining a constant baseline velocity and amplitude of pulsing. The POD technique \citep{aubry91,Berkooz1993} has been commonly used to describe and elucidate large-scale spatially orthogonal coherent structures in single-phase flows. In fact, in a recent study analogous to the work presented, \citet{higham2018implications} discusses the suitability of POD in quasi-two-dimensional (Q2D) liquid flows. We use the energy budget of the spectral modes and their interactions to describe nonlinearities in the system, leading to the observed bubbling patterns. The paper is organized as follows: the experimental setup is described at the beginning, followed by a discussion on periodic time-averaging and POD. Finally, we conclude by presenting the results from flow-field reconstruction using the dominant modes based on POD analysis.

\section{Experimental setup} \label{Experiments}
\begin{figure}[h!]
\centering
\includegraphics[width=6cm]{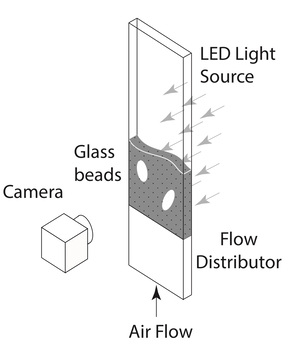}
\caption{Illustration of the experimental setup (not to scale).}
\label{fig:setup}
\end{figure}

An experimental investigation was undertaken at the National Energy Technology Laboratories, Department of Energy, Morgantown, USA. A custom built, Q2D bed with acrylic windows 50mm wide, 5mm deep, and 300mm tall (Fig.~\ref{fig:setup}) was used. The chosen dimensions conform to a Q2D flow, where the resulting meso-scale structures (bubbles) are larger than the shortest dimension. 18g of 400$\mu$m glass beads were added to the system resulting in a 50mm tall bed. A uniform flow was created using a fractal flow distributor. The gas flow rate at the inlet is pulsed in the form of a sine wave given by:
\begin{equation}
    Q(t) = A + Bsin(2f \pi t)
\end{equation}

\noindent where, A=2.6 l/min and B=2.1 l/min represent the baseline and amplitude of the pulsed wave form. Three different frequencies, viz. 4Hz, 5Hz and 6Hz. The images were captured using a 120mm Nikon lens and Fastec IL5L sensor at 300Hz. The bed was lit from the back using an LED light source such that the high-intensity bubble patterns could be recorded using a threshold procedure. Also, the front of the test section was lit to track individual glass particles using PTVResearch (\citep{PTVResearch}). The solid-phase velocity field was obtained by binning individual particle tracks using a cubic interpolator. The PODDEM algorithm of \citet{higham2016rapid} was implemented to detect the outliers and construct the field accordingly.

\begin{figure}
    \centering
    \includegraphics[width=\textwidth]{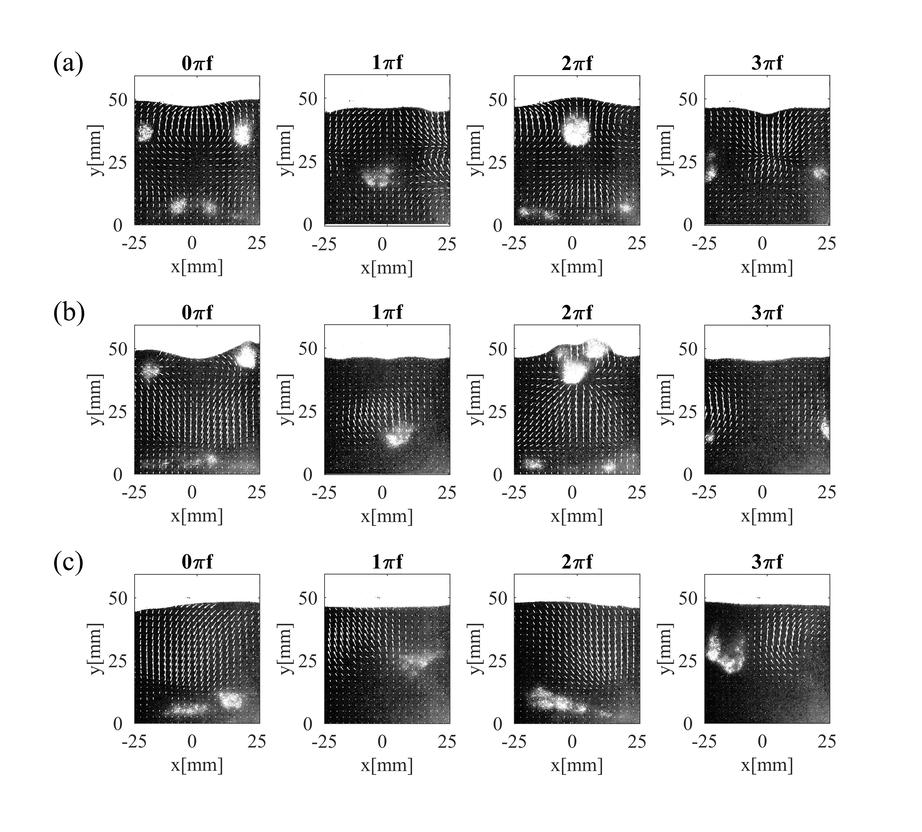}
    \caption{Snapshots of bubbles patterns with velocity vectors overlaid for f = 6Hz (a), 5Hz (b) \& 4Hz (c).}
    \label{fig:bubble}
\end{figure}

Instantaneous snapshots from the experiments show repeatable bubbling patterns. At 6Hz and 5Hz frequencies, a recurring structure having two bubbles along the sides followed by a single bubble along the centre is produced (see Fig.~\ref{fig:bubble}a\&b). We refer to this as a one-two (1:2) pattern. when the frequency of the wave form is reduced to 4Hz, the location of bubbles alternates from one side to another over two complete pulse cycles and we refer to this as a one-one (1:1) pattern (Fig.\ref{fig:bubble}c). The bubble size is observed to increase as the operating frequency is reduced. The findings are consistent with the work of \citet{massimilla1966study}, who showed increasing the frequency of the inlet gas decreases the size of the bubble and changes the pattern. As highlighted by the quiver plots in Figs.~\ref{fig:bubble} (a-c), the resulting bubble characteristics is modifying the granular dynamics.



\section{Analysis}
\subsection{Periodic time-averaging} \label{TimeAvg}
\begin{figure}
	\includegraphics[width=\textwidth]{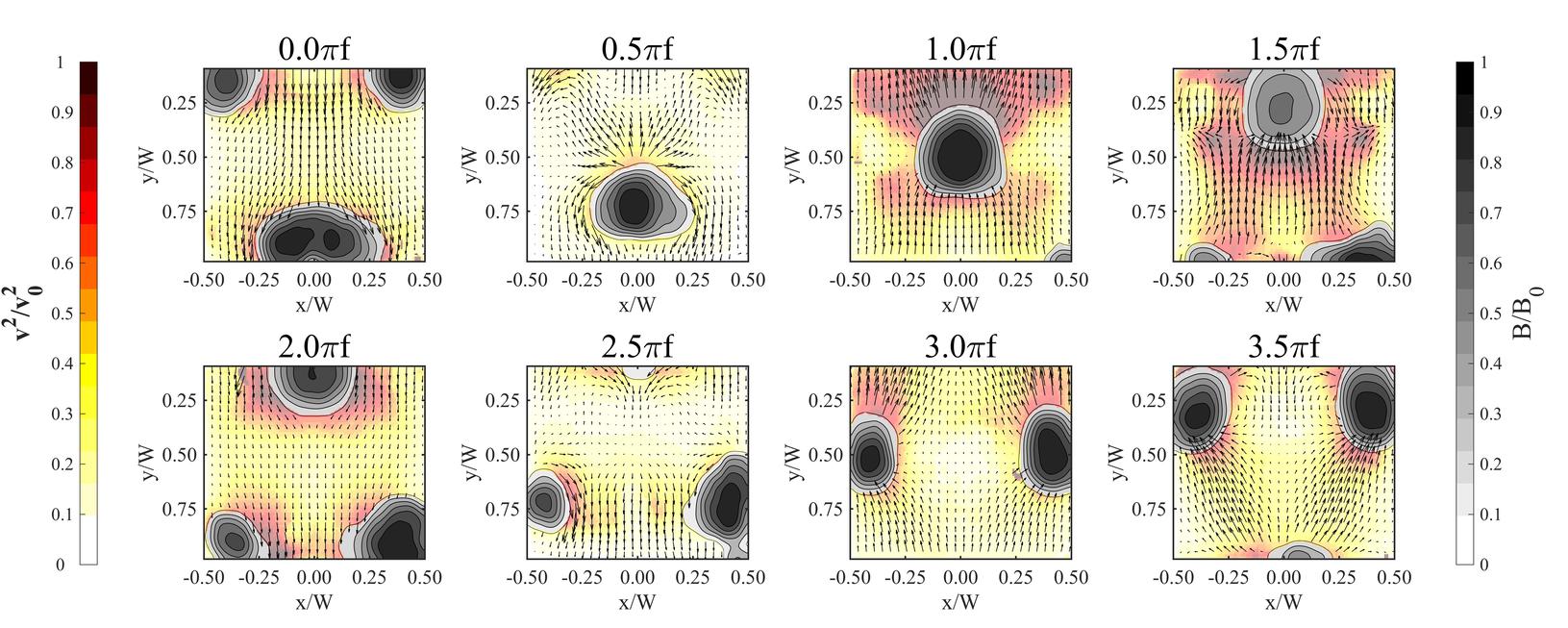}
	\caption{Periodic time-averaged bubble patterns and velocity fields at different phase angles for the 6Hz input frequency.}  
	\label{fig:all_recon_case_6hz}
\end{figure}

\begin{figure}
	\includegraphics[width=\textwidth]{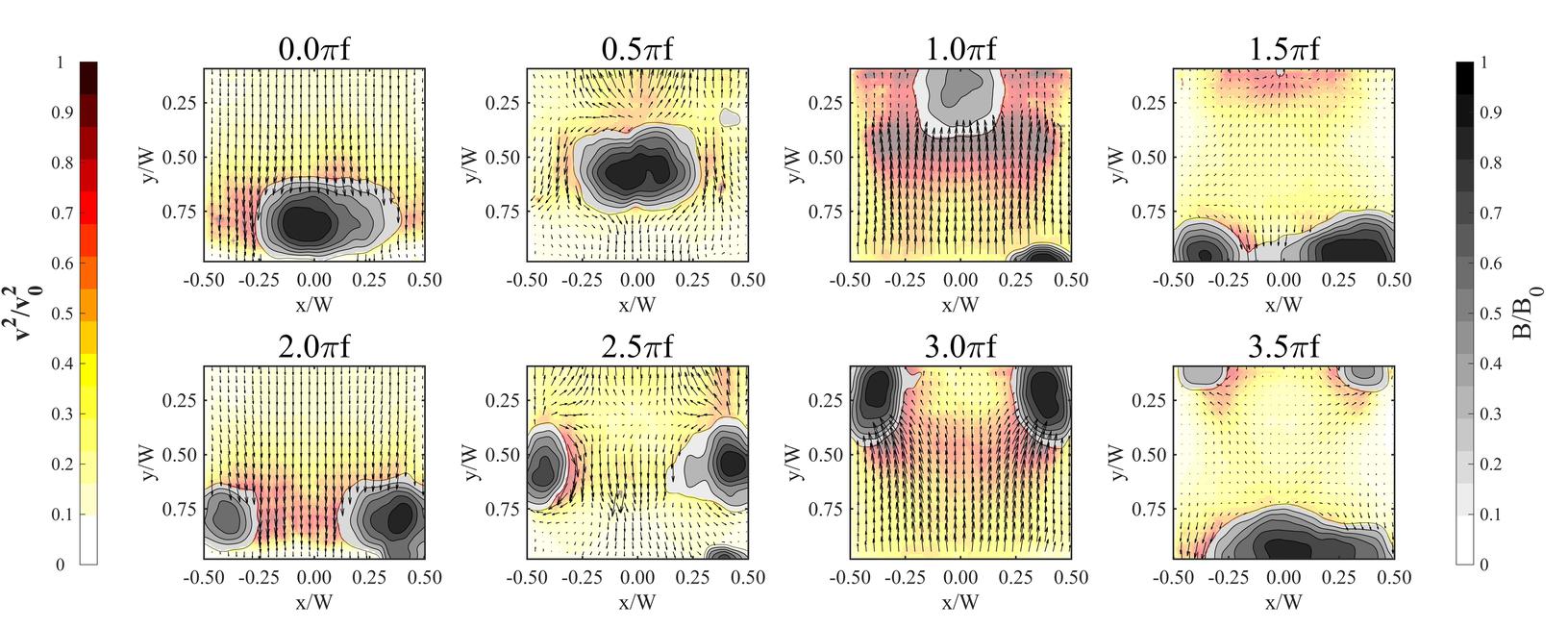}
	\caption{Periodic time-averaged bubble patterns and velocity fields at different phase angles for the 5Hz input frequency.}  
	\label{fig:all_recon_case_5hz}
\end{figure}

\begin{figure}
	\includegraphics[width=\textwidth]{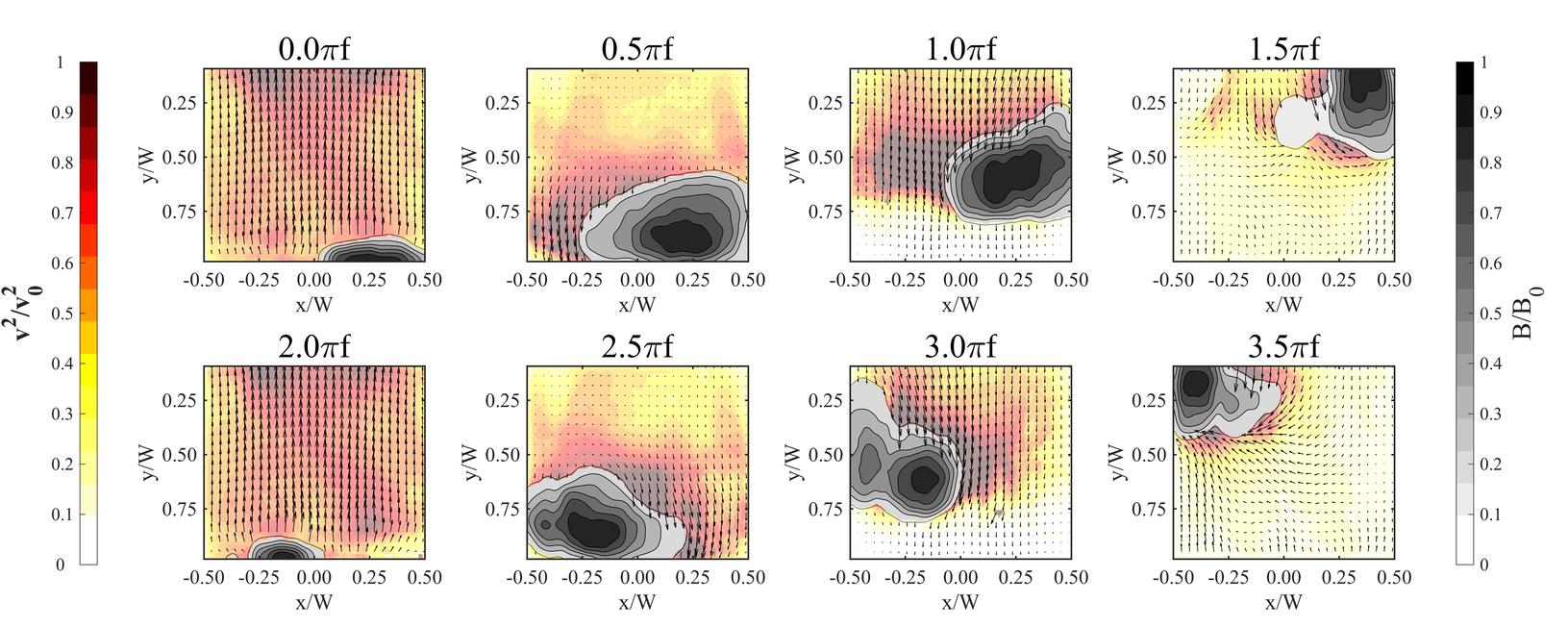}
	\caption{Periodic time-averaged bubble patterns and velocity fields at different phase angles for the 4Hz input frequency.}
	\label{fig:all_recon_case_4hz}
\end{figure}

Periodic time-averaging is obtained by averaging the data at fixed time intervals. As shown in Section \ref{Experiments}, the patterns repeat over two periods of pulsing frequency. Therefore the interval used between data for averaging is two periods of the inlet frequency along which eight linearly spaced temporal locations are chosen. The statistics are generated from 100 samples in each case to deter any bias. As part of this investigation, two independent calculations are performed: First the average location of bubbles is determined. The bubbles identified by the thresholding procedure are binarised, where the voids are defined by ones and the solid phase by zeros. By periodically time-averaging the bubble data, two-dimensional histograms of bubble locations ($\mathbf{B}$) are created, which are normalized by the maximum value ($\mathbf{B}_0$). Second, the time-averaged sum of the square of fluctuating velocity components ($\mathbf{v}^2$) is determined which represents granular temperature of the solid-phase. These values are normalized using the square of the maximum inlet gas velocity ($\mathbf{v}_0^2$). The bubble location and granular temperature distribution are shown in Figs.\ref{fig:all_recon_case_6hz},~\ref{fig:all_recon_case_5hz} \&~\ref{fig:all_recon_case_4hz}. In each figure the bubble histograms are represented in grey, and granular temperature in color, with quiver plots of velocity components overlaid. The results indicate that as the frequency of the inlet gas is decreased, the bubbles become larger and their locations (represented by histograms) become less predictable similar to the findings of \citet{massimilla1966study}. 


\subsection{Proper Orthogonal Decomposition} \label{POD}

POD is a statistical method commonly used to extract and analyse spatial coherence. POD technique was independently developed by several researchers, and is consequently referred to in different areas of applications as Karhunen-Lo\`{e}ve Decomposition, Singular Value Decomposition (SVD) and Principal Components Analysis (PCA) \citep{kosambi, loeve, Karhunen, Pougachev, obukov}. POD extracts the most relevant modes from a set of stochastic signals. POD can be thought of as fitting an n-dimensional ellipsoid to the data, where each axis is represented by spatially orthogonal eigenvectors computed using a co-variance matrix built from the data.

It is common in fluid dynamics to decompose the flow field spectrally by applying POD on a set of snapshots \citep{Sirovich1987}. The data set is comprised of instantaneous field variables obtained from computer simulation or experiment. The decomposition results in a set of dominant modes that represent an average spatial description of structures predominantly associated with large-scale motion. The modes may also correspond to events in flow that contribute the most to its energy in a statistical sense. This methodology has been successfully used on several occasions to explain complex flow phenomenon in fluid dynamics \citep{Aubry1988,Rowley2004,Sirovich1987,higham2018modification,higham2017using}. 

The POD analysis based on the method of snapshots is summarised as follows: A set of $t = 1, 2, \ldots,T$ temporally ordered snapshots, $\mathbf{V}(x,y;t)$, is considered, each of which is of dimension $X \times Y$. The method requires constructing an $N \times T$ matrix $\mathbf{W}$ from  $T$  columns, of length $N = XY$, each corresponding to a column-vector version of a  transformed snapshot $\mathbf{V}(x,y;t)$. A POD is obtained using the singular value decomposition formulated as:
\begin{equation}
\mathbf{W} \equiv \mathbf{\Phi} \, \mathbf{S} \, \mathbf{C}^{*}  
\label{eq:svd}
\end{equation}
where, the matrix $\mathbf{\Phi}$ of dimension ${\Theta \times \Theta}$ contains the spatial structure, and the matrix $\mathbf{C}$ of dimension ${\Theta \times \Theta}$ pertains to the temporal evolution of each mode. $\Theta$ represents the number of modes in the truncated decomposition. The $(\cdot)^*$ represents the conjugate transpose matrix operation. The matrix $\mathbf{S}$ of dimension ${\Theta \times \Theta}$ contains the singular values of matrix $\mathbf{W}$. A vector  ${\lambda} = \text{diag}(\mathbf{S})^{2} / {(N-1)}$ can then be constructed from $\mathbf{S}$ containing the contribution of each mode to the total variance. The elements in $\lambda$ are ordered in descending rank order, i.e. ($\lambda_{1} \ge \lambda_{2} \ge \ldots \lambda_{\Theta} \ge 0$). If the modes are computed from the fluctuating velocity fields, $\lambda_i$s are associated with granular temperature in each mode and their relative contribution can be defined by:
\begin{equation}
E(\%) = \frac{\lambda_i}{\sum^N_{i=1} \lambda_i} \cdot 100
\end{equation}

\begin{figure}
    \centering
    \includegraphics[width=\textwidth]{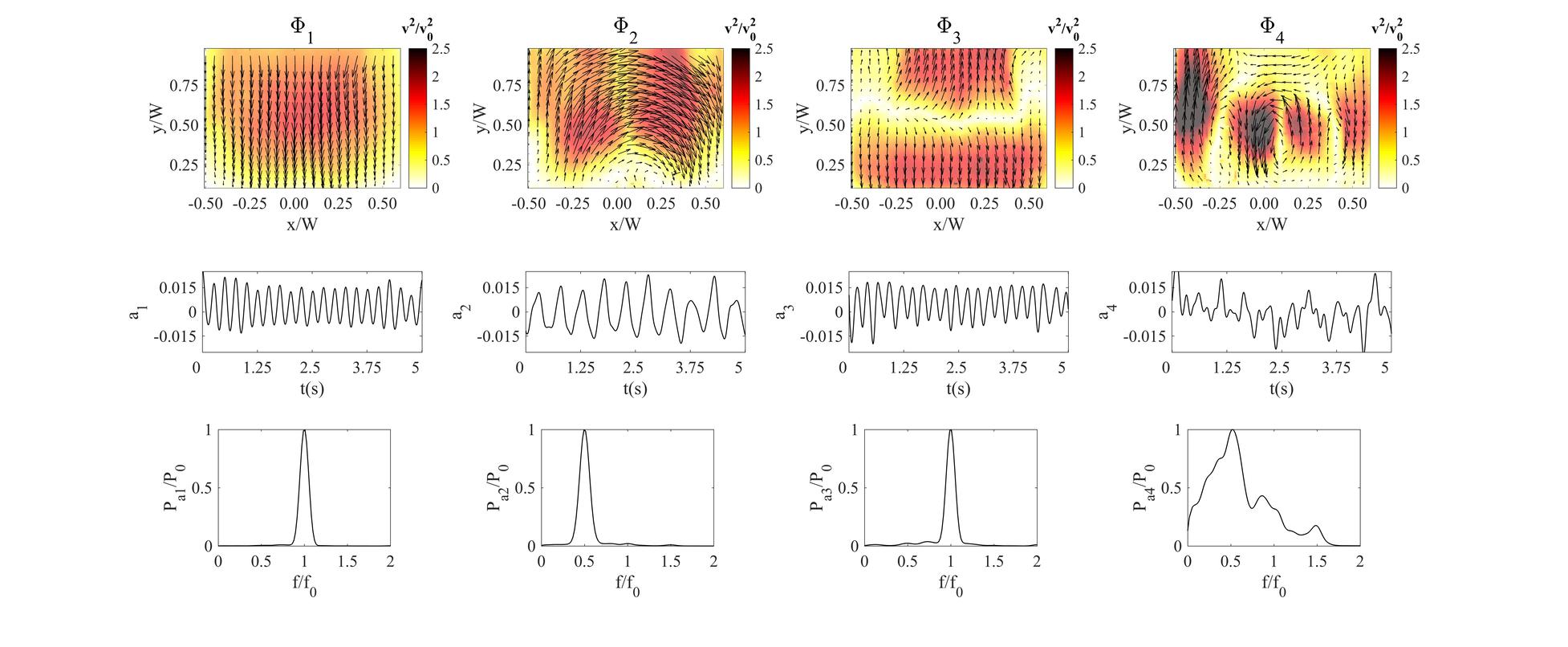}
    \caption{POD of 4Hz case: The top row shows the leading spatial modes $\mathbf{\Phi}$, here the components are plotted as the sum of the square of velocity fluctuations divided by the maximum inlet gas velocity ($\mathbf{v}/v_0$) with overlaid quiver plots. The middle row shows their temporal coefficients $\mathbf{C}_i$, and the bottom row shows the Fourier transform of temporal coefficients.}
    \label{fig:pod4}
\end{figure}

\begin{figure}
    \centering
    \includegraphics[width=\textwidth]{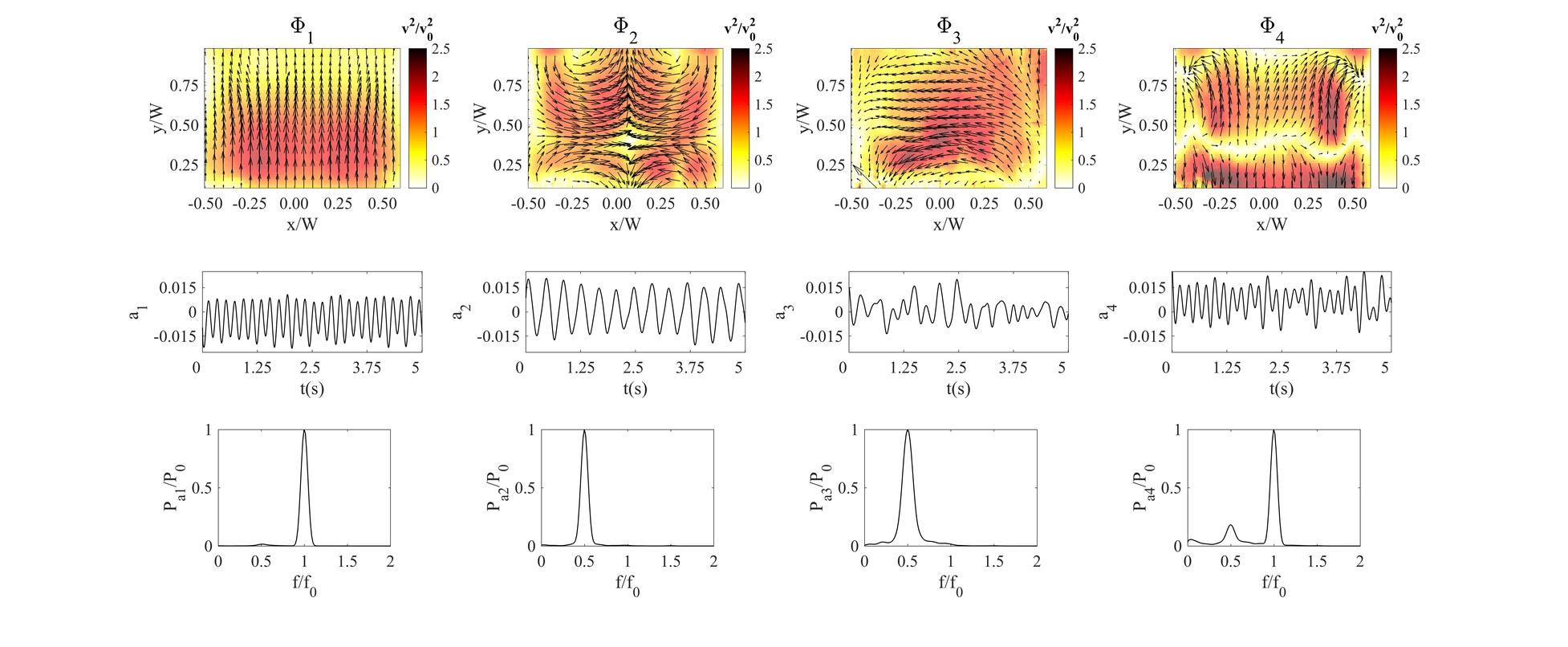}
    \caption{POD of 5Hz case: The top row shows the leading spatial modes $\mathbf{\Phi}$, here the components are plotted as the sum of the square of velocity fluctuations divided by the maximum inlet gas velocity ($\mathbf{v}/v_0$) with overlaid quiver plots. The middle row shows their temporal coefficients $\mathbf{C}_i$, and the bottom row shows the Fourier transform of temporal coefficients.}
    \label{fig:pod5}
\end{figure}

\begin{figure}
    \centering
    \includegraphics[width=\textwidth]{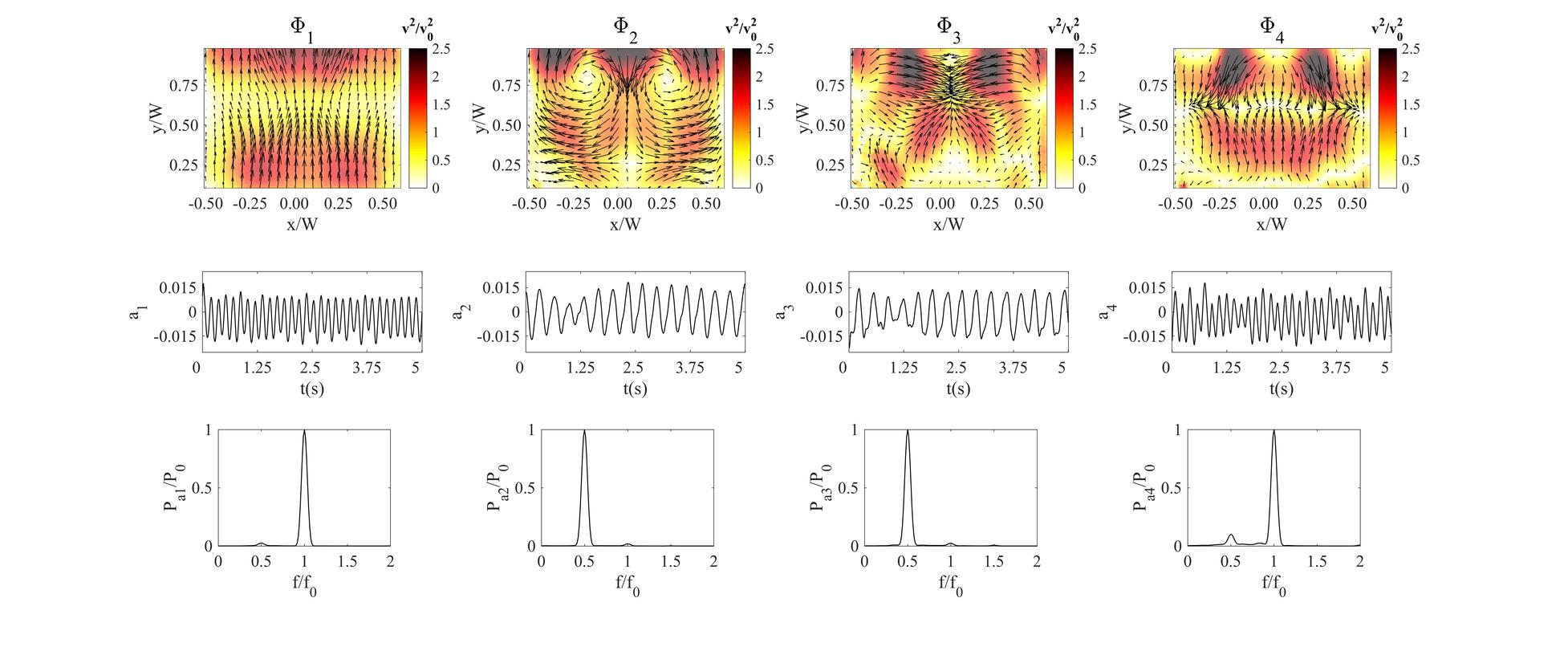}
    \caption{POD of 6Hz case: The top row shows the leading spatial modes $\mathbf{\Phi}$, here the components are plotted as the sum of the square of velocity fluctuations divided by the maximum inlet gas velocity ($\mathbf{v}/v_0$) with overlaid quiver plots. The middle row shows their temporal coefficients $\mathbf{C}_i$, and the bottom row shows the Fourier transform of temporal coefficients.}
    \label{fig:pod6}
\end{figure}

\begin{figure}
    \centering
    \includegraphics[width = 6cm]{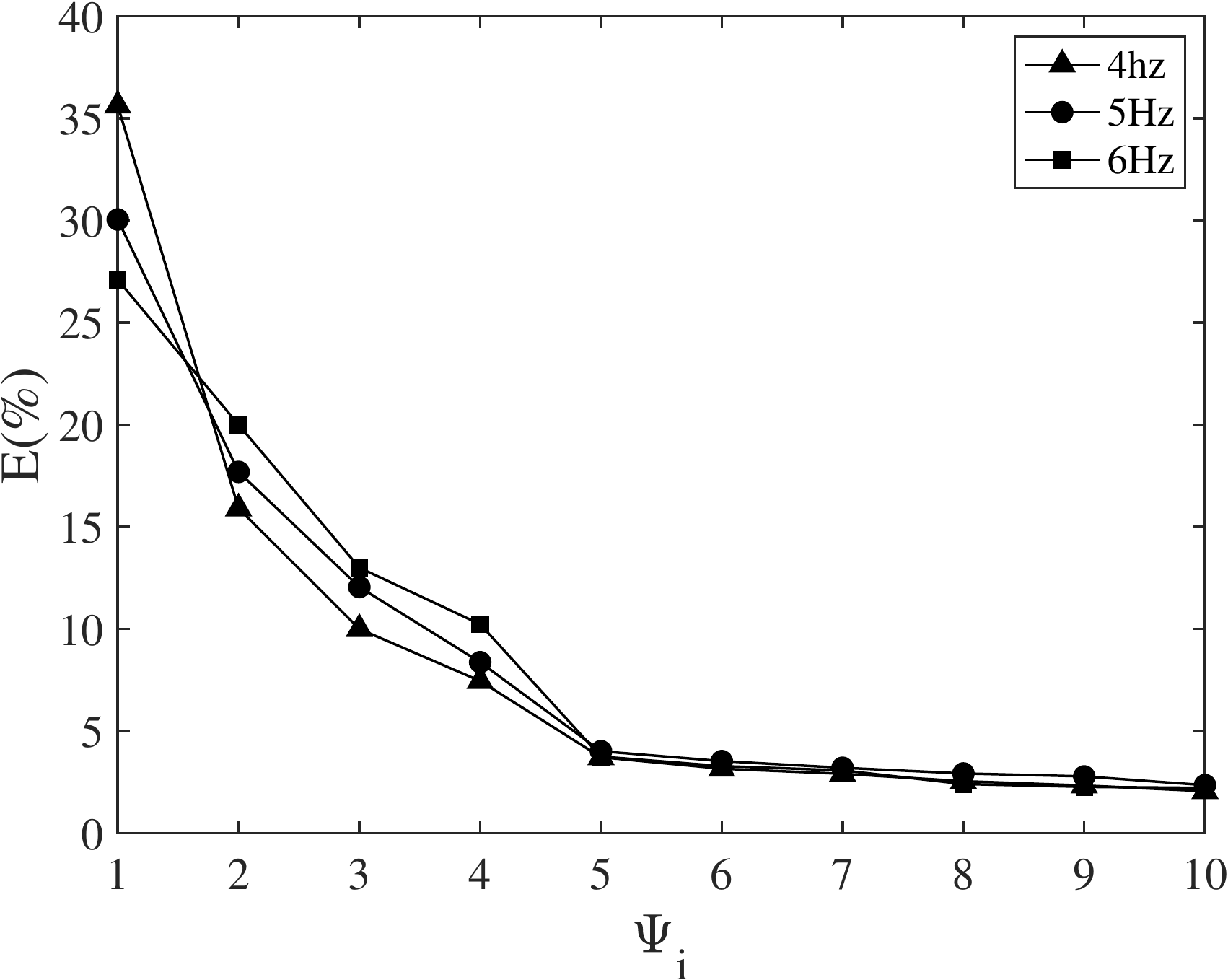}
    \caption{Relative energy contribution (E\%) of each POD mode}
    \label{fig:pod_spec}
\end{figure}

In the current analysis, we use POD for a modal decomposition of granular dynamics in the pulsed-fluidised bed system described in Section \ref{Experiments}. The POD computations are applied to the fluctuating solids velocity field and in each case 15,000 snapshots are used. The results from the first four modes are presented in Figs.~\ref{fig:pod4}, \ref{fig:pod5} \& \ref{fig:pod6}. Only the first four leading modes are plotted and these account for $\sim$ 70\% of the total energy (Fig.~\ref{fig:pod_spec}). The spatial modes $\mathbf{\Phi_{i}}$s are shown on the top row of each POD vizualisation. The middle row shows the temporal coefficients $\mathbf{C}_i$s, which map the spatial modes through time, and the bottom row shows the Fourier spectra of temporal coefficients. 

It is possible to make a number of observations. In all cases, the POD modes are dominated by two frequencies, harmonic (the pulsing frequency) and sub-harmonic (half of the pulsing frequency). Spatially they relate to two separate dynamics: The harmonic frequency is associated with stream-wise transport of particles and the sub-harmonic frequency is related to their transverse motion. In all the cases, $\mathbf{\Phi}_{1}$, the mode with the largest energy contribution relates to the harmonic forcing and $\mathbf{\Phi}_{2}$ relates to the sub-harmonic frequency. The change in pulsing frequency clearly has an effect on modal decomposition of energy as shown in Fig.~\ref{fig:pod_spec}. For instance, as the frequency is decreased, the contribution of $\mathbf{\Phi}_{1}$ increases (from 27\% to 30\% to 35\%) while the contribution of $\mathbf{\Phi}_{2}$ decreases (from 20\% to 17\% to 15\%). This trend continues for higher-order modes $\mathbf{\Phi}_{3}$ \& $\mathbf{\Phi}_{4}$. 

In 6Hz and 5Hz cases, $\mathbf{\Phi}_{3}$ relates to the sub-harmonic frequency, and $\mathbf{\Phi}_{4}$ to the harmonic frequency. There is a noticeable secondary peak associated with $\mathbf{\Phi}_{4}$ whose amplitude increases from 6Hz to 5Hz. In the 4Hz case, $\mathbf{\Phi}_{3}$ is associated with the harmonic frequency and resembles a conjugate pair of $\mathbf{\Phi}_{1}$, although only accounting for 15\% of the total energy. The Fourier power spectrum of $\mathbf{\Phi}_{4}$ has a dominant peak at the sub-harmonic frequency, however the spectrum is colluded with multiple peaks unlike 6Hz and 5Hz cases. 

An important trend which emerges from the POD analysis is the redistribution of energy between the harmonic and sub-harmonic components. For 6Hz and 5Hz cases, the harmonic ($\mathbf{\Phi}_{1\&4}$) and sub-harmonic modes ($\mathbf{\Phi}_{2\&3}$) account for 38\% \& 32\% of the total energy contribution. In comparison for the 4Hz case, the relative contributions from the harmonic ($\mathbf{\Phi}_{1\&3}$) and sub-harmonic ($\mathbf{\Phi}_{2\&4}$) modes are 45\% \& 25\%. Hence, there is an evident shift in energy among the dominant frequency components. The energy content in the harmonic modes related to the vertical motion is transferred to the sub-harmonic modes associated with the lateral motion as the pulsing frequency is increased. This leads to a higher degree of isotropy in the system, which might eventually result in an efficient mixing of solids.


\subsection{Flow field reconstruction}

\begin{figure}
	\includegraphics[width=\textwidth]{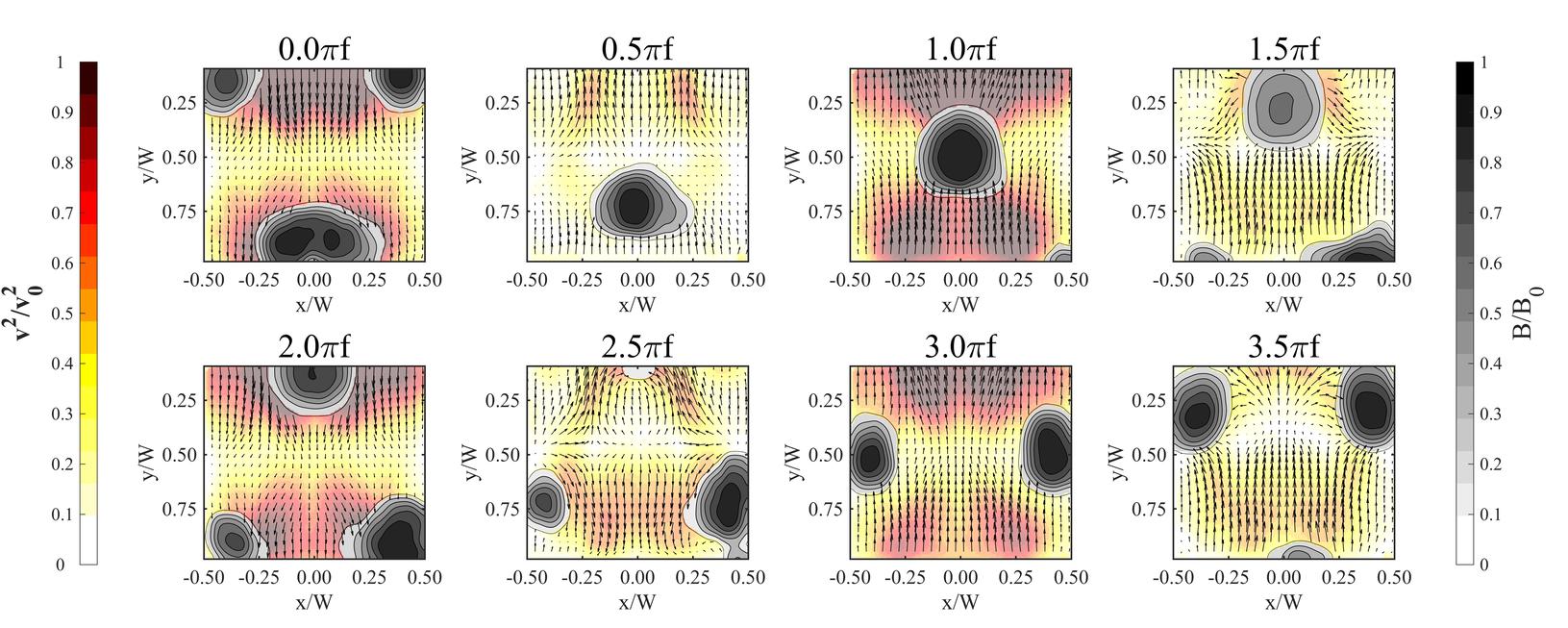}
	\caption{Reconstruction of the flow-field using $\mathbf{\Phi}_{1\& 4}$ representing the harmonic frequency for the 6Hz case.}
	\label{fig:r1-6}
\end{figure}

\begin{figure}
	\includegraphics[width=\textwidth]{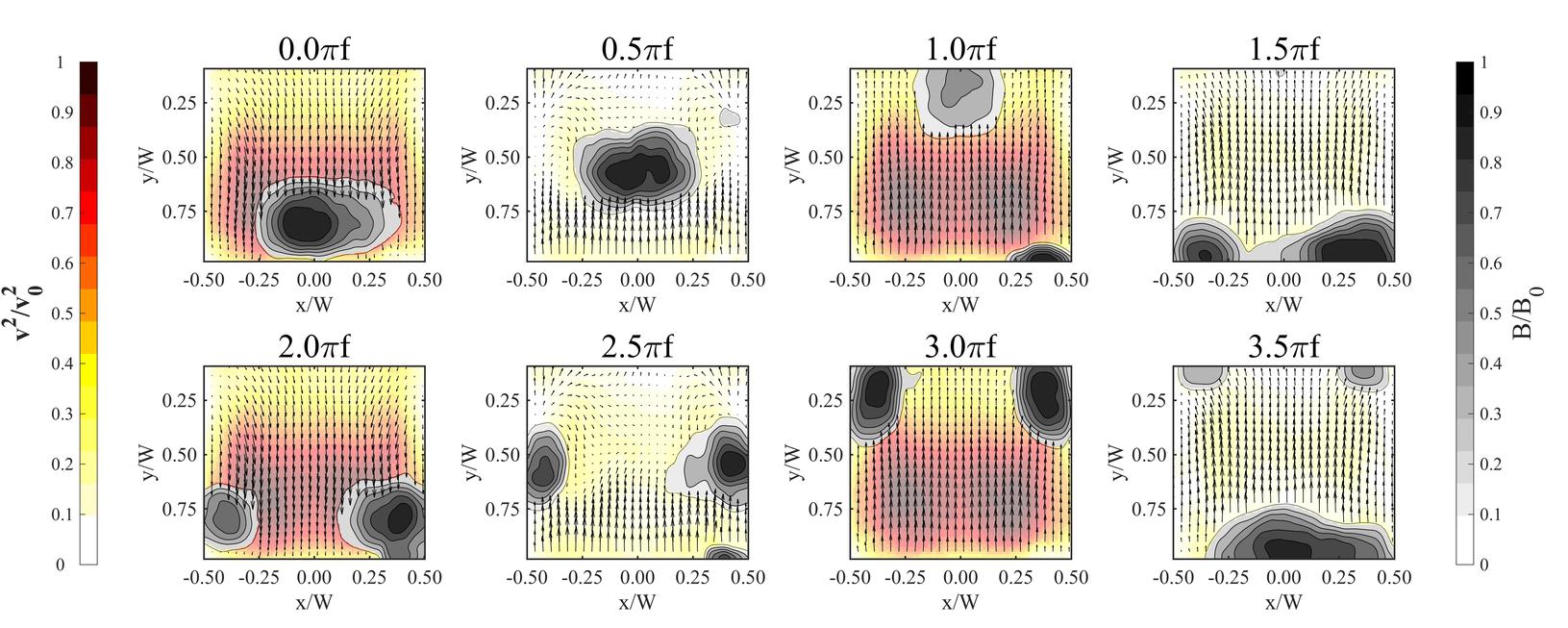}
	\caption{Reconstruction of the flow-field using $\mathbf{\Phi}_{1\& 4}$ representing the harmonic frequency for the 5Hz case.}
	\label{fig:r1-5}
\end{figure}

\begin{figure}
	\includegraphics[width=\textwidth]{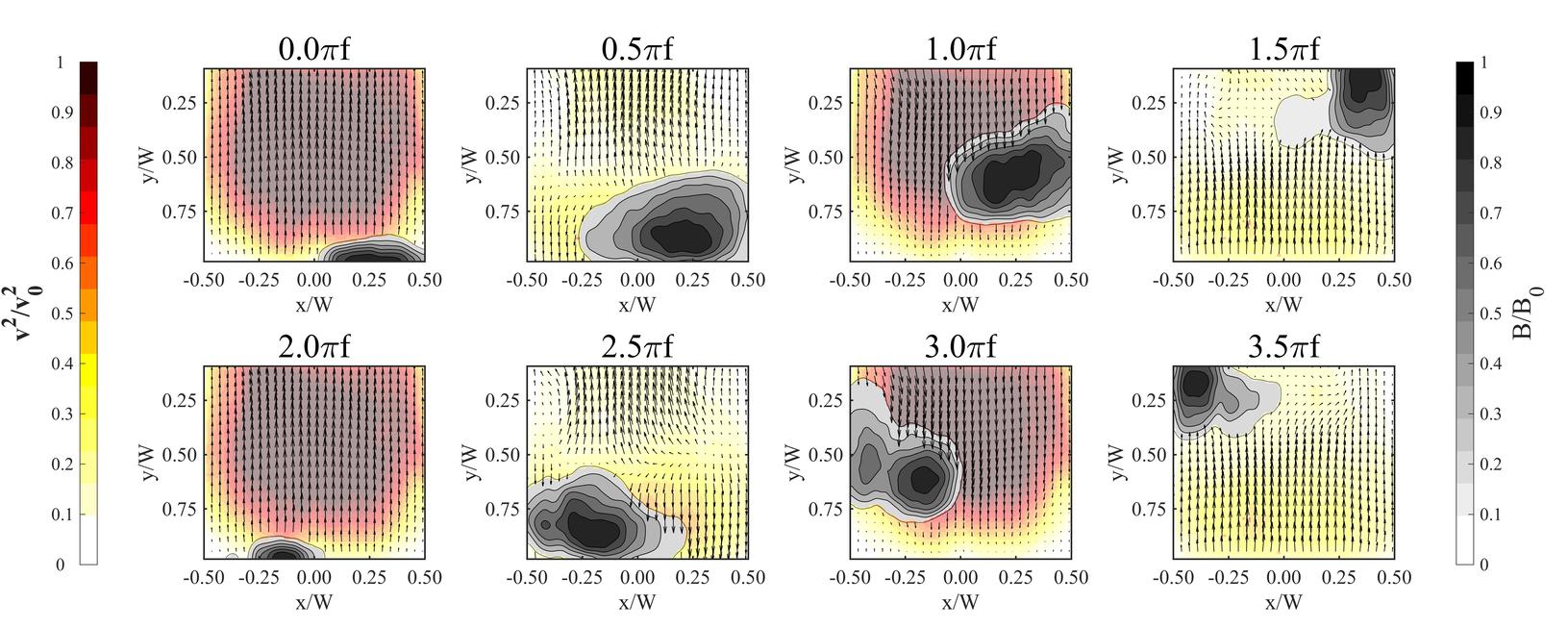}
	\caption{Reconstruction of the flow-field using $\mathbf{\Phi}_{1\& 3}$ representing the harmonic frequency for the 4Hz case.}
	\label{fig:r1-4}
\end{figure}

Using the POD modes, singular values and temporal coefficients, it is possible to make a low-order reconstruction of fluctuating velocity fields (Eq.~\ref{eq:svd}) followed by periodic time-averaging similar to Section \ref{TimeAvg}. Figs.~\ref{fig:r1-6}, \ref{fig:r1-5}~\&~\ref{fig:r1-4} show the flow fields reconstructed from modes associated with the harmonic frequency, superimposed with the two-dimensional bubble histograms. Figs.~\ref{fig:r1-6} \& ~\ref{fig:r1-5} are created from modes one and four contributing to $\sim$38\% and $\sim$40\% of the total energy for 6Hz and 5Hz frequencies. For the 4Hz case, the first and third modes representing $\sim$45\% of the total energy are used (Fig.~\ref{fig:r1-4}). The vector plots confirm that the inlet frequency relates to vertical momentum transfer due to bubbles, in line with the results in Section \ref{POD}.

\begin{figure}
	\includegraphics[width=\textwidth]{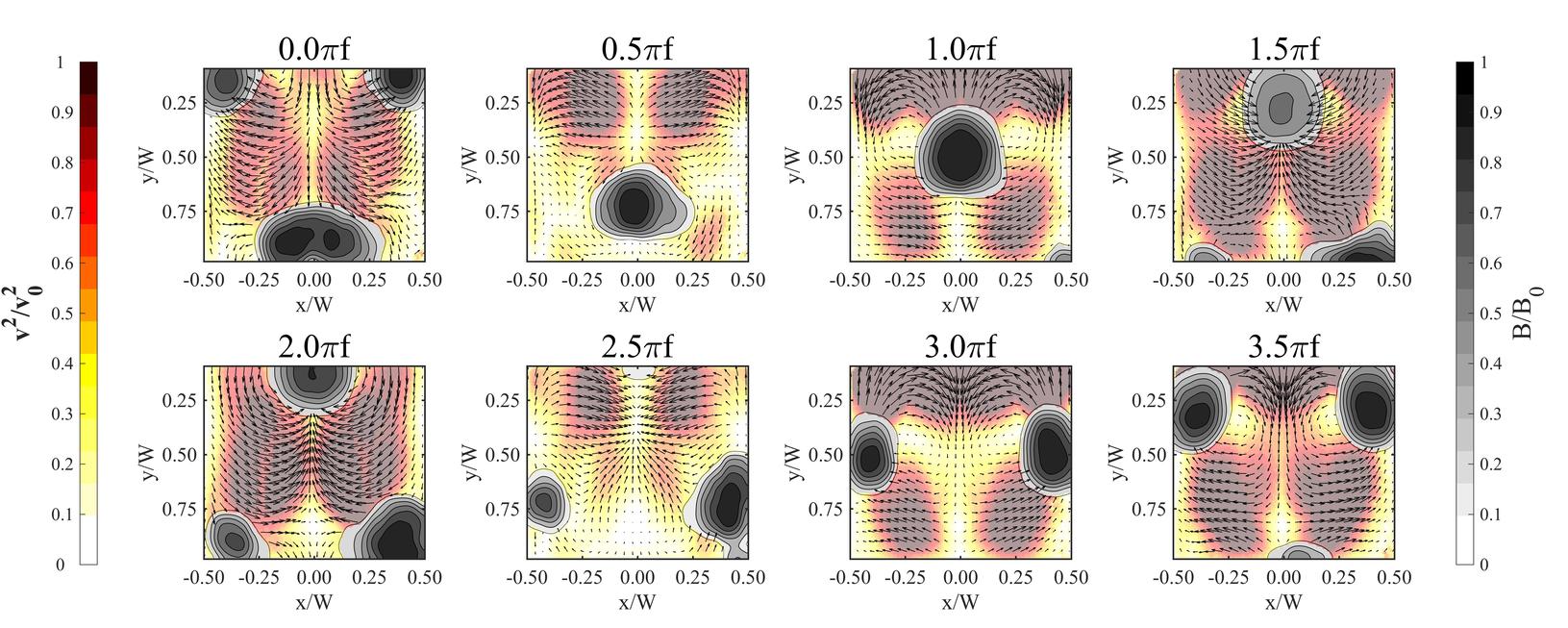}
	\caption{Reconstruction of the flow-field using $\mathbf{\Phi}_{2\& 3}$ representing the sub-harmonic frequency for the 6Hz case.}
	\label{fig:rp5-6}
\end{figure}

\begin{figure}
	\includegraphics[width=\textwidth]{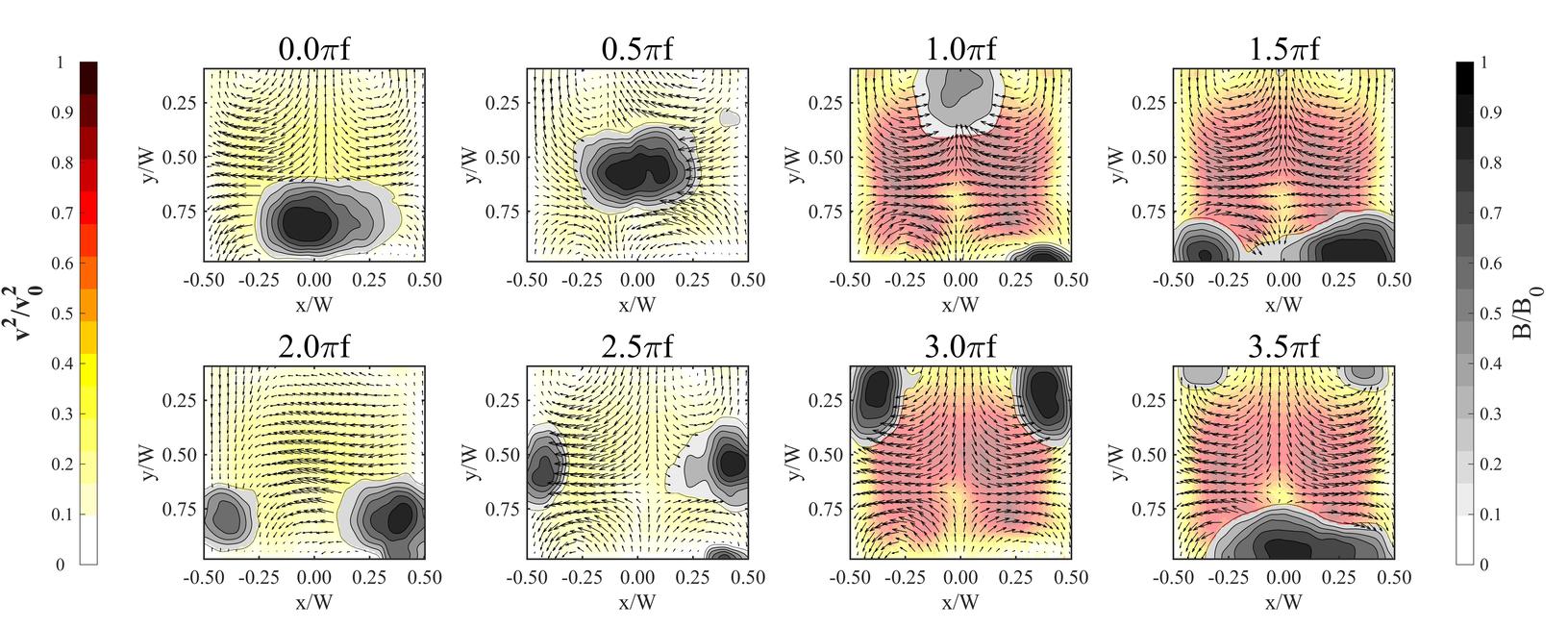}
	\caption{Reconstruction of the flow-field using $\mathbf{\Phi}_{2\& 3}$ representing the sub-harmonic frequency for the 5Hz case.}
	\label{fig:rp5-5}
\end{figure}

\begin{figure}
	\includegraphics[width=\textwidth]{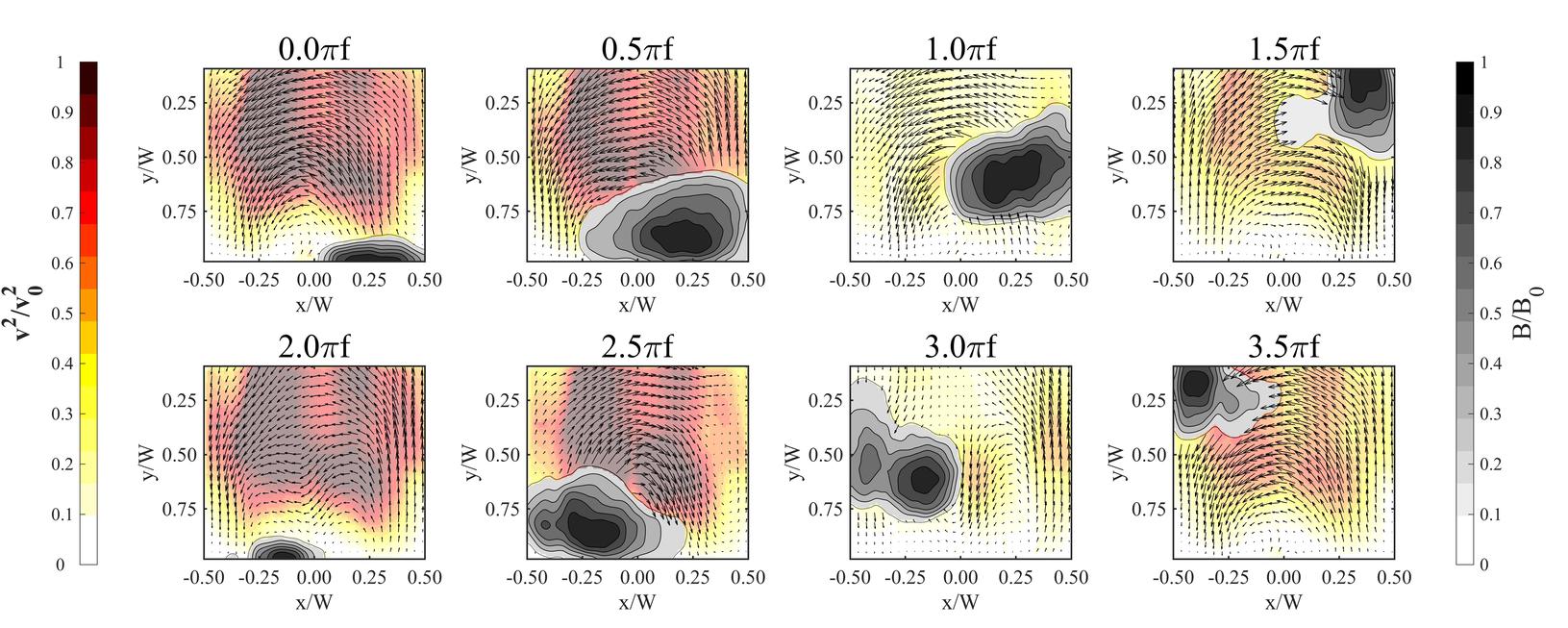}
	\caption{Reconstruction of the flow-field using $\mathbf{\Phi}_{2\& 4}$ representing the sub-harmonic frequency for the 4Hz case.}
	\label{fig:rp5-4}
\end{figure}

\begin{figure}
	\includegraphics[width=\textwidth]{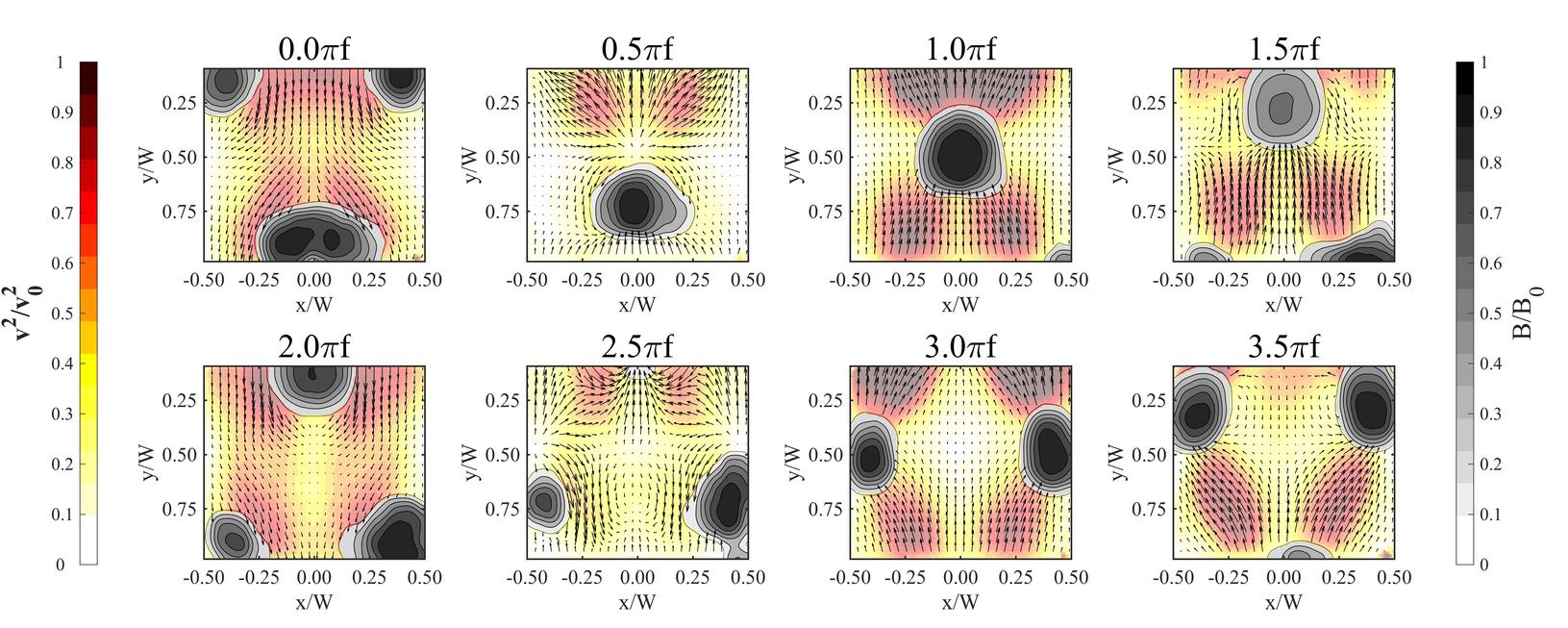}
	\caption{Reconstruction of the flow-field using $\mathbf{\Phi}_{1-4}$ for the 6Hz case.}
	\label{fig:r14-6}
\end{figure}

\begin{figure}
	\includegraphics[width=\textwidth]{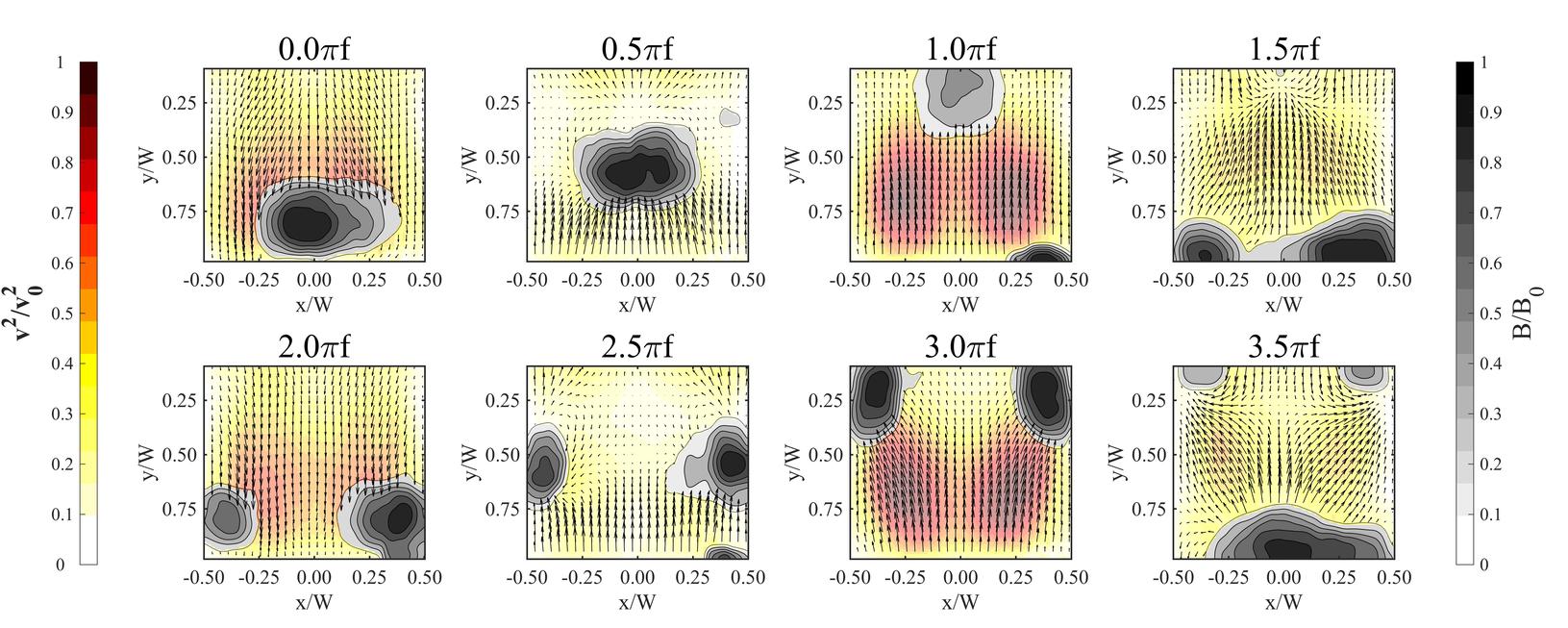}
	\caption{Reconstruction of the flow-field using $\mathbf{\Phi}_{1-4}$ for the 5Hz case.}
	\label{fig:r14-5}
\end{figure}

\begin{figure}
	\includegraphics[width=\textwidth]{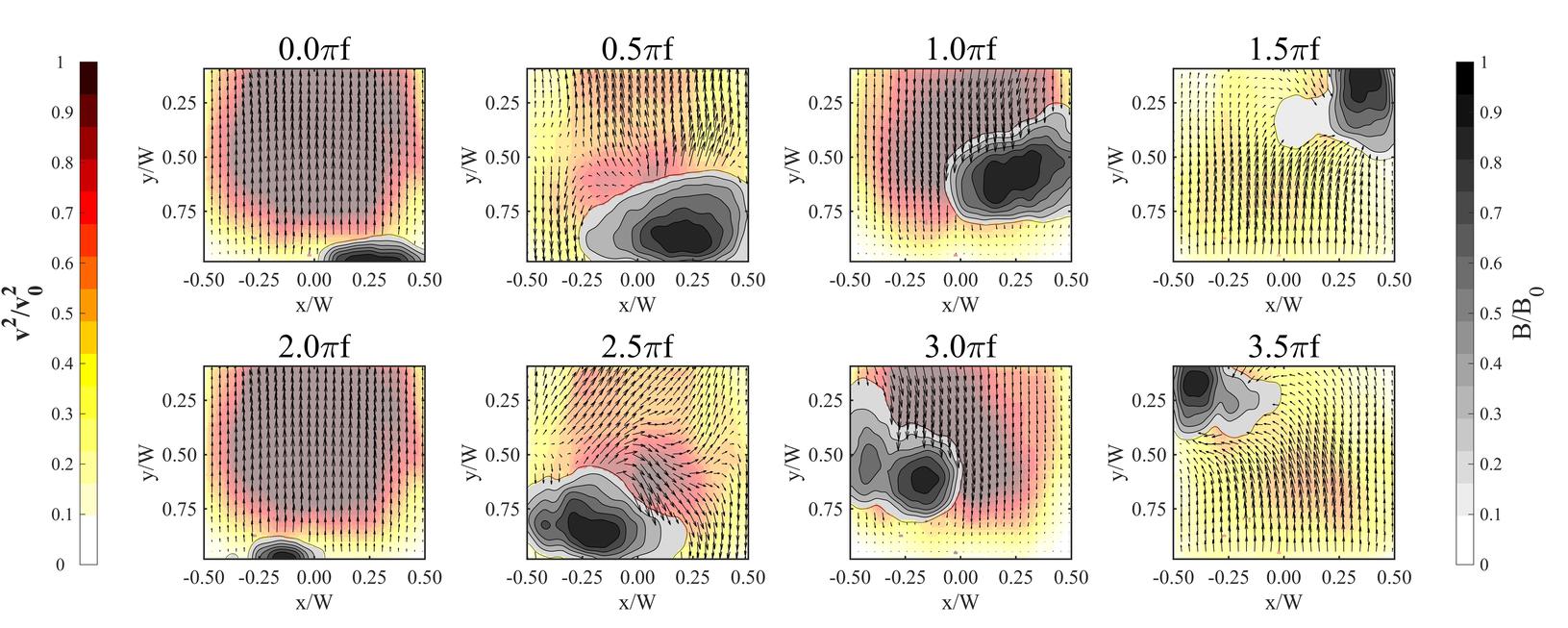}
	\caption{Reconstruction of the flow-field using $\mathbf{\Phi}_{1-4}$ for the 4Hz case.}
	\label{fig:r14-4}
\end{figure}

Likewise, the flow field reconstructed using sub-harmonic modes are shown in Figs. ~\ref{fig:rp5-6}, ~\ref{fig:rp5-5} \& ~\ref{fig:rp5-4}. Figs.~\ref{fig:rp5-6} \& ~\ref{fig:rp5-5} are created from $\mathbf{\Phi}_{2\&3}$ contributing to $\sim$34\% and $\sim$ 29\% of the total energy for the 6Hz and 5Hz cases respectively. Fig.~\ref{fig:rp5-4} is created using $\mathbf{\Phi}_{2\&4}$ which represents $\sim$21\% of the total energy when the frequency is 4Hz. It is apparent that these modes relate to the mixing process inside the bed, created by shearing as the bubbles pass through the solids. This could be verified by vortical structures in the flow fields reconstructed using the sub-harmonic components. The size these structures increases while their frequency of occurrence decreases as the the inlet frequency is decreased. Finally, we use the four leading POD modes to create a low-order representation of flow fields at different operating frequencies as shown in Figs.\ref{fig:r14-6}, \ref{fig:r14-5}~\&~\ref{fig:r14-4}. The results are encouraging in that they reproduce the results from periodic time-averaging of raw data with reasonable accuracy. The quality of reconstruction improved appreciably at higher frequency of pulsing which could be due to a greater suppression of chaos as pointed out earlier.

Based on the results presented, the bubbling mechanism in the pulsed-fluidized bed could be explained as follows: There is a continuous interchange of solids between the wake region and the surrounding granular medium along the trajectory of a bubble \citep{Davidson1966}. Initially as the bubbles are produced at the inlet, there is an energy intense region around the bubble. During the initial stages of its ascent, the wake is characterised by a low-energy region. As the bubble continues to travel through the bed, its entrainment results in a high-energy region in its wake. The size and number of bubbles characterise the wake structure. For the 6Hz and 5Hz cases, as the leading bubble passes through the centre of the bed, its wake entrains particles resulting in a high-energy region downstream. The continuous displacement of particles generates two identical paths of least resistance on either side of the wake. As the leading bubble reaches the surface of the bed, two distinct bubbles are produced along these paths and this process repeats to generate the 1:2 pattern. From the POD results and reconstructions it is clear the development of this patterns relates to the energy contribution of either the inlet frequency causing vertical motion, or its sub-harmonic component related to mixing. For the 4Hz case, there is a change in bubbling pattern, the large alternating single bubbles follow the walls of the bed. This would cause entrainment to occur at one end of the domain. Consequently, as it reaches the surface of the bed, there is a path of least resistance for a bubble on the other end due to the displacement of particles being entrained by the leading bubble. 


\section{Conclusions}
In the present study, we have used POD analysis to characterize granular dynamics in a Q2D pulsed-fluidized bed. We observe that the flow fields are dictated by the harmonic and sub-harmonic frequencies. These relate to stream-wise transport and transverse motion of particles respectively. Using POD, the energy is decomposed in a spectral sense, which could then be correlated to the observed behavior. The bubbling pattern changes (from 1:2 to 1:1) as the frequency is decreased from 6Hz to 4Hz, while maintaining a constant baseline velocity and amplitude of pulsing. There is a corresponding redistribution of energy between the harmonic and sub-harmonic modes. The particle-velocity field becomes more isotropic at higher frequencies which could lead to a better mixing of solids. This is also accompanied by a greater suppression of chaos, confirmed by the resulting bubble location histograms. Furthermore, we have used low-order modes from POD analysis to reconstruct the flow-field. There is a greater resemblance with the time-averaged raw data for 6Hz and 5Hz cases compared to the 4Hz case which could again point to the suppression of non-linear dynamics at higher operating frequencies.

\section{Acknowledgements}
The work has been supported by the Oak Ridge Institute of Science and Education (ORISE). The authors are grateful for the support and guidance provided by the National Energy Technology Laboratory, Department of Energy (NETL-DOE). The authors are also particularly grateful to the MFAL team for technical support and to Dr. Bill Rogers for providing technical insight, resources and facilities.

\bibliographystyle{jphysicsB}
\bibliography{bib}
\end{document}